\documentclass[aps,twocolumn,a4paper,pra,nofootinbib]{revtex4}
\usepackage{psfig}
\usepackage{amssymb}
\usepackage{latexsym}
\usepackage{amsfonts}

\begin{document}

\title{Recent Experiments on the Casimir Effect: Description and Analysis}
\author{Astrid Lambrecht}
\author{Serge Reynaud}
\email{reynaud@spectro.jussieu.fr}
\homepage{www.spectro.jussieu.fr/Vacuum}
\affiliation{Laboratoire Kastler Brossel, case 74, 
Campus Jussieu, 75252 Paris, France}
\thanks{Laboratoire du CNRS, de l'ENS 
et de l'Universit{\'e} Pierre et Marie Curie} 

\author{(S\'eminaire Poincar\'e, 2002)}

\maketitle

\section{Motivations}

After its prediction in 1948 \cite{Casimir48}, the Casimir force has been observed 
in a number of `historic' experiments which confirmed its existence and main 
properties \cite{Sparnaay89,Milonni94,Mostepanenko97,LamoreauxResource99}. 
The Casimir force has recently been measured with a 
largely improved experimental precision \cite{Bordag01} which allows for an 
accurate comparison between measured values of the force and theoretical predictions. 
This comparison is interesting for various reasons. 

The Casimir force is the most accessible effect of vacuum fluctuations 
in the macroscopic world. As the existence of vacuum energy raises difficulties 
at the interface between the theories of quantum and gravitational phenomena, 
it is worth testing this effect with the greatest care and highest accuracy 
\cite{Reynaud01,Genet02Iap}. But the comparison between theory and experiment 
should take into account the important differences between the real experimental 
conditions and the ideal situation considered by Casimir. 

Casimir calculated the force between a pair of perfectly smooth, flat and 
parallel plates in the limit of zero temperature and perfect reflection. 
He found an expression for the force $F_{\rm Cas}$ 
and the corresponding energy $E_{\rm Cas}$ which only depend 
on the distance $L$, the area $A$ and two fundamental constants, the speed of light 
$c$ and Planck constant $\hbar$
\begin{eqnarray} 
F_{\rm Cas} &=& \frac{\hbar c \pi ^2 A}{240L^4} = -
\frac{{\rm d} E_{\rm Cas}}{{\rm d}L} \nonumber \\
E_{\rm Cas} &=&  \frac{\hbar c \pi^2 A}{720 L^3}  
\label{Fcasimir} 
\end{eqnarray}
Each transverse dimension of the plates has been supposed to be much larger 
than $L$. Conventions of sign are chosen so that $F_{\rm Cas}$
and $E_{\rm Cas}$ are positive. They correspond to an attractive force
($\sim 0.1 \mu$N for $A=1 {\rm cm}^2$ and $L=1\mu$m) 
and a binding energy.

Most experiments have been performed in a sphere-plane geometry
which differs from the plane-plane geometry considered by Casimir. 
In the former geometry, the force
is derived from the Deriagin approximation \cite{Deriagin68}, often 
called in a somewhat improper manner the proximity force theorem. 
With this approximation, the force is obtained as the integral of force 
contributions corresponding to the various inter-plate distances as if 
these contributions were independent. 
In the plane-sphere geometry, the force is thus determined by the
radius $R$ of the sphere and by the Casimir energy as evaluated in the 
plane-plane configuration. 
The Deriagin approximation is also used to evaluate the surface 
roughness corrections. 

The fact that the Casimir force (\ref{Fcasimir}) only depends on fundamental
constants and geometrical features is remarkable. In particular it is independent 
of the fine structure constant which appears in the expression of the atomic
Van der Waals forces. This universality property is related to the assumption
of perfect reflection used by Casimir in his derivation. Perfect mirrors 
correspond to a saturated response to the fields since they reflect 100 \% 
of the incoming light. This explains why the Casimir effect, though it has its 
microscopic origin in the interaction of electrons with electromagnetic fields, 
does not depend on the fine structure constant. 
Now, real mirrors are not perfect reflectors.
The most precise experiments are performed with metallic mirrors which 
behave as nearly perfect reflectors at frequencies smaller than a characteristic
plasma frequency but become poor reflectors at higher frequencies. Hence
the Casimir expression has to be modified to account for the effect of finite
conductivity. At the same time, experiments are performed at room temperature 
whereas the Casimir formula (\ref{Fcasimir}) only holds in vacuum, that is at
zero temperature. 

A precise knowledge of the Casimir force is a key point in many accurate 
force measurements for distances ranging from nanometer to millimeter. 
These experiments are motivated either by tests of Newtonian gravity at 
millimetric distances \cite{Fischbach98,Hoyle01,Adelberger02,Long02} 
or by searches for new weak forces predicted 
in theoretical unification models with nanometric to millimetric ranges 
\cite{Carugno97,Bordag99,Fischbach99,Long99,Fischbach01}.
Basically, they aim at putting limits on deviations of experimental results 
from present standard theory. 
The Casimir force is the dominant force between two neutral 
non-magnetic objects in the range of interest so that any new force would
appear as a difference between experimental measurements and theoretical 
expectations of the Casimir force.

As far as the aim of a theory-experiment comparison is concerned, the accuracy of 
theory is as crucial as the precision of experiments. If a given accuracy, say at 
the $1 \%$ level, is aimed at in the comparison, then the theoretical and 
experimental accuracy have to be mastered at this level independently from each 
other. Since the various corrections to the Casimir formula which have already 
been alluded to may have a magnitude much larger than the 1\% level, a high-accuracy 
comparison necessarily requires a precise analysis of the differences between the 
ideal case considered by Casimir and real situations studied in experiments. 

\section{Experiments before 1997}

We first review some of the experiments performed before 1997.

The first experiment to measure the Casimir force between two metals was carried 
out by Spaarnay in 1958 \cite{Sparnaay58}. 
A force balance based on a spring balance was used to measure 
the force between two flat neutral plates for distances between 0.5 and 
2$\mu$m. Measurements were carried out for Al-Al, Cr-Cr and Cr-steel plates 
through electromechanical techniques. Spaarnay discussed the major difficulties
of the experiments, in particular the control of the parallelism of the two 
plates, the determination of the distance between them, and the control of 
the neutrality of the two metal plates which is delicate since the Casimir force
can easily be masked by electrostatic forces. 
The experiment gave evidence of an attractive force between the two plates and 
Sparnaay cautiously reported that ``the observed attractions do not contradict 
Casimir's theoretical prediction''.
For the sake of comparison with experiments described below,
an error bar of the order of 100 \% may be attributed to this experiment.

Probably the first unambiguous measurement of the Casimir force between metallic 
surfaces was performed by van Blokland and Overbeek in 1978 \cite{Blokland78}. 
The force was measured with the help of a spring balance between a lens and a flat 
plate, both coated with 50-100nm thick chromium layers, for distances from 132 
to 670nm, measured by determining the capacitance of the system. 
The use of a lens instead of a second flat plate simplified the control 
of the geometry by suppressing the problem of parallelism.
The force in this configuration was evaluated with the help of 
Deriagin's approximation discussed in more detail below.
The investigators compared their experimental results to theoretical
calculations using the Lifshitz theory for chromium and concluded to 
an agreement between the measured and calculated force values, 
confirming for the first time the effect of finite conductivity.
For this experiment, one may estimate the accuracy to be of the order 
of 25\%.

The Casimir force has been observed in a number of other experiments,
in particular \cite{Deriagin57,Tabor68,Black68,Sabisky73}.
More detailed or systematic reviews may be found in 
\cite{Sparnaay89,Milonni94,Mostepanenko97,LamoreauxResource99,Bordag01}. 

\section{Recent experiments}
Recently new measurement techniques were used to measure the Casimir effect with 
improved accuracy. Quite a number of experiments have been carried out in the last 
years and we will describe some of them which seem to be the most
significant ones. 

In 1997 Steve Lamoreaux measured the Casimir force by using a torsion pendulum 
\cite{Lamoreaux97}. The force was measured between a metallized sphere and a 
flat metallic plate with controlled but unequal electrostatic potential. 
Since the electrostatic and Casimir forces were acting simultaneously, it was  
necessary to substract precisely the effect of the electrostatic force 
in order to deduce the value of the Casimir force. This measurement 
was made for distances between 0.6 and 6 microns. Comparison between 
the experimental results and the theoretical predictions was 
reported to be in agreement at the level of 5 \%.

After the correction of inaccuracies in the initial report
\cite{Lamoreaux98e,Lambrecht00prl,Lamoreaux98r}, the results of this
experiment can be summarized as follows~: the force has
been measured, probably with an error bar of the order of 10 \% at the 
shortest distances where the effect of finite conductivity of the Au and
Cu metallic layers used in the experiments was unambiguously observed;
the error bar was certainly much larger at distances larger than a few $\mu$m
where the magnitude of the force is much weaker; this probably explains why 
the temperature correction has not been seen though it should have been seen 
at the largest distance $\sim 6 \mu$m explored in the experiment (see below).
It is difficult to be more affirmative on this topic, in particular
because this experiment was stopped by the relocation of Steve Lamoreaux.

\begin{figure}[htb]
\centerline{\psfig{figure=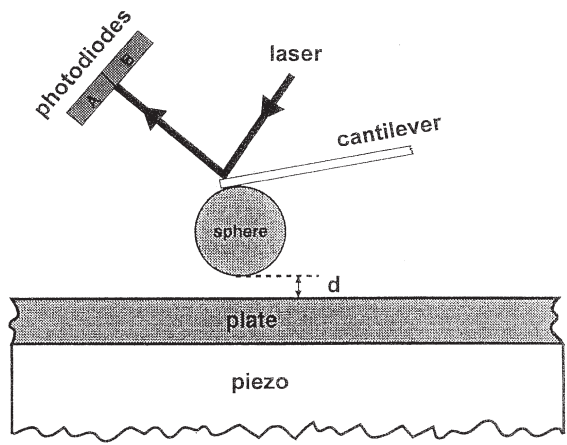,width=8cm}}
\caption{Experimental setup of the Casimir force measurement 
in \cite{Mohideen98,Roy99pr,Harris00}. The force is measured between the
sphere and the plate with the distance of closest approach d (denoted 
$L$ in the present report). The sphere is fixed on the cantilever of an AFM 
and its position measured by the deflection of a laser beam on the 
top of the cantilever. With kind courtesy of Umar Mohideen.}
\label{mohideenexp}
\end{figure}

Shortly after this publication, a second measurement was reported by 
Umar Mohideen \cite{Mohideen98} followed by several reports 
with an improved precision \cite{Roy99pr,Harris00}. 
This experiment is based on the use of an atomic force microscope (AFM). 
A metallized sphere is fixed on the cantilever of the microscope and brought 
to the close vicinity of a flat metallic plate, at a distance between 
0.1 and 0.9$\mu$m. Both surfaces are put at the same electrostatic potential
and the Casimir force is measured by the deflection of a laser beam on the 
top of the cantilever, as shown on Figure \ref{mohideenexp}.

\begin{figure}[htb]
\centerline{\psfig{figure=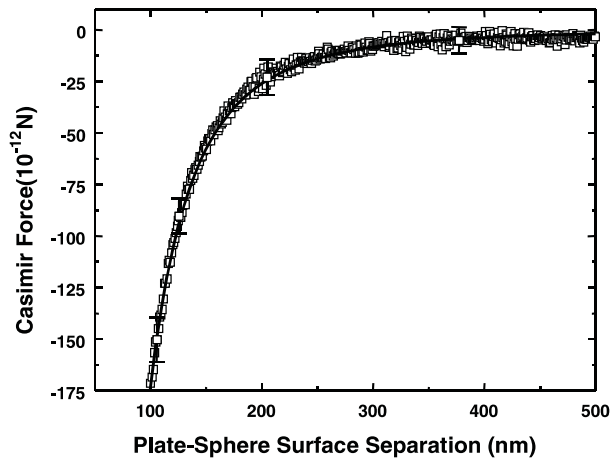,width=8cm}}
\caption{Comparison between experimentally measured values and theoretical 
predictions of the Casimir force, as reported in \cite{Roy99pr}; 
the squares and bars represent experimental points and errors bars for
a few of them; the solid line represents theoretical predictions.
With kind courtesy of Umar Mohideen.}
\label{mohideenres}
\end{figure}

The comparison between experimental results and theoretical predictions 
has been performed for Al and Au coated surfaces. A typical experimental accuracy
at the level of 1\% is obtained with a comparable agreement with theory,
as depicted on Figure \ref{mohideenres}.
Theoretical points are based on the methods described below~:
they take into account the large effects of conductivity and 
plane-sphere geometry as well as the effects of temperature
and roughness which have a smaller influence ($<$1\%) for this experiment.
The same group has also studied the effect of sinusoidal corrugations
on the properties of the Casimir force \cite{Roy99} and has observed 
the lateral Casimir force between corrugated plates \cite{Chen02}.

An independent measurement has been published in 2000 by Thomas Ederth
\cite{Ederth00} who also used an AFM. 
The force was measured between two neutral metallic
crossed cylinders (curvature 10mm) at short distances ranging from 20 to 100nm. 
Great efforts allowed Ederth to reduce the surface roughness, which was
a necessity at these short distances. After a careful 
error analysis, Ederth concluded to an accuracy at the level of several \%.

The experiments done by Federico Capasso and his group at Bell Labs,
using microelectromechanical systems (MEMS), also deserve a special mention
\cite{Chan01s,Chan01p}. MEMS are movable structures fabricated on a semiconductor 
wafer through integrated circuit technology and they are used as a new generation 
of sensors and actuators working in the micrometer or submicrometer distance range.
The Casimir force is measured between a polystyrene sphere and a polysilicon plate
with metallic coatings. The plate is suspended so that it can 
rotate around an axis. The variation of the plate rotation angle when the sphere
is approached to a distance between 100nm and 1$\mu$m reveals the Casimir force
with a magnitude agreeing with theory. 
When the plate is set into oscillation, frequency shifts, histeretic behavior
and bistability are observed, again in agreement with the effect of the Casimir
force predicted by the theory.
The main interest of these experiments is to show that the Casimir force 
plays a significant role in systems of technological interest like the MEMS. 
It is indeed the dominant force in the micrometer range
and this experiment shows that mechanical effects of quantum vacuum fluctuations 
have to be taken into account in micro- or nanotechnology \cite{Bishop01}.  

Experiments described in the present section up to this point use a sphere-plane
geometry or a crossed cylinders geometry. Their analysis relies on the
accuracy of the Deriagin approximation which is not precisely known.  
This is not the case for the experiments performed in the initial Casimir
geometry with two parallel flat plates.  
A measurement in this geometry has recently been reported on by Bressi,
Carugno, Onofrio and Ruoso \cite{Bressi01,Bressi02}. 
The force is observed between two parallel flat plates coated with chromium, 
one of which is mounted on a silicon cantilever while the other one is fixed
on a rigid piezoelectric stack. The plate fixed on the piezoelectric stack
is set into oscillatory motion and this induces a varying Casimir force
onto the plate mounted on the cantilever. The motion of the latter is then
monitored by using a tunneling electromechanical transducer. 
The measurement has been performed for distances between 0.5 and 3$\mu$m 
and the result has been found to agree with theory at the 15\% precision level.

\section{The effect of imperfect reflection}

As explained in the introduction, a precise theory-experiment comparison 
requires not only a detailed control of the experiments but also 
a careful estimation of the theoretical expectation of the force in the
real conditions of the experiments. We begin here by the more spectacular
``correction'' to the ideal Casimir formula (\ref{Fcasimir}) which is
associated with imperfect reflection of mirrors.

No real mirror can be considered as a perfect reflector at all field 
frequencies. In particular, the most precise experiments are performed with
metallic mirrors which show perfect reflection only at frequencies smaller 
than a characteristic plasma frequency $\omega_{\rm P}$ which depends on the
properties of conduction electrons in the metal. Hence the Casimir force 
between metal plates does fit the ideal Casimir formula (\ref{Fcasimir}) only 
at distances $L$ much larger than the plasma wavelength  
\begin{equation}
\lambda _{\rm P}=\frac{2\pi c}{\omega _{\rm P}}
\end{equation}
For metals used in the recent experiments, this wavelength lies in the 
$0.1\mu$m range (107nm for Al and 136nm for Cu and Au).
At distances smaller than or of the order of the plasma wavelength, the finite 
conductivity of the metal has a significant effect on the force.
The idea has been known since a long time \cite{Lifshitz56,Schwinger78}
but the investigation of the effect of imperfect reflection
has been systematically developed only recently.

We first consider the initial Casimir geometry with perfectly plane, flat and 
parallel plates at zero temperature. We thus restrict our attention on the effect
of the reflection properties of the mirrors described by scattering 
amplitudes which depend on the frequency of the incoming field. Assuming that these
amplitudes obey general properties of unitarity, high-frequency transparency and 
causality, one derives a regular expression of Casimir force which is free from the 
divergences usually associated with the infiniteness of vacuum energy.
The cavity formed by the two mirrors can be dealt with by using the Fabry-P{\'e}rot 
theory. Vacuum field fluctuations impinging the cavity have their energy either 
enhanced or decreased inside the cavity, depending on whether their frequency is 
resonant or not with a cavity mode. The radiation pressure associated with these 
fluctuations exerts a force on the mirrors which is directed either inwards or 
outwards respectively. It is the balance between the inward and outward 
contributions, when they are integrated over the field frequencies and 
incidence angles, which gives the net Casimir force \cite{Jaekel91}. 

The techniques of analytical continuation of the response functions
in the complex plane then allow one to write the Casimir force as an 
integral over imaginary frequencies $\omega = i \xi$ with $\xi$ real
\begin{eqnarray}
F &=&\frac{\hbar A}{\pi }\sum_{p} \int \frac{{\rm d}^{2}{\bf k}}{4\pi ^{2}}
\int\limits_{0}^{\infty} {\rm d}\xi 
\frac{\kappa r_1^p \left[i\xi,{\bf k}\right] r_2^p \left[i\xi,{\bf k}\right]}
{e^{2\kappa L} - r_1^p \left[i\xi,{\bf k}\right] r_2^p \left[i\xi,{\bf k}\right]}  
\nonumber \\
&&\kappa \equiv \sqrt{{\bf k}^2 +\frac{\xi^2}{c^2}}  \label{RealMirrors}
\end{eqnarray}
$r_j^p \left[\omega,{\bf k}\right]$ is the reflection amplitude for the two mirrors
$j=1,2$ and the field mode characterized by a frequency $\omega$, a tranverse
wavevector ${\bf k}$ (transverse means orthogonal to the main direction of the 
cavity, that is also parallel to the plane of the plates) and a polarization $p$.
The amplitudes appear in expression (\ref{RealMirrors}) at imaginary frequencies
$\omega = i \xi$ where they have real and positive values. 
The fraction appearing in (\ref{RealMirrors}) represents the difference
between the radiation pressures on outer and inner sides of the cavity 
after the continuation to the imaginary axis. It is determined by
the product of the reflection amplitudes of the two mirrors and by
an exponential factor $e^{2\kappa L}$ representing the propagation 
dephasing for the field after a roundtrip in the cavity, that is
a propagation length $2L$.
Expression (\ref{RealMirrors}) includes the contribution of the modes
freely propagating inside and outside the cavity but also the contribution
of evanescent waves confined to the vicinity of the mirrors. 

Equation (\ref{RealMirrors}) is a convergent integral for any couple of 
mirrors described by scattering amplitudes obeying the properties of causality, 
passivity and high frequency transparency. This means that the potential divergence  
associated with the infiniteness of vacuum energy has been cured by using the
physical properties of scattering amplitudes, that is also by describing mirrors 
just as opticians do.
Furthermore expression (\ref{RealMirrors}) does not depend on any particular 
microscopic model but may be applied to any reflection amplitude obeying the
general properties already discussed. 
Itcan be used for lossy as well as lossless mirrors \cite{Genet02prep}. 

The ideal Casimir result is recovered at the limit where mirrors may be considered 
as perfect over the frequency range of interest, that is essentially over the first 
few resonance frequencies of the cavity \cite{Jaekel91}.
This can be considered as an alternative demonstration of the Casimir formula 
without any reference to a renormalization or regularization technique.
For real mirrors, the effect of imperfect reflection is described by a reduction factor 
$\eta _{\rm F}$ which multiplies the ideal Casimir expression (\ref{Fcasimir}) 
to give the force $F$ 
\begin{eqnarray}
F &=& \eta _{\rm F} F_{\rm Cas}  
\label{defEta}
\end{eqnarray}

In order to go further, we have to specialize the general expression (\ref{RealMirrors}) 
to a model of mirrors. The commonly used model corresponds to reflection on bulk mirrors 
with an optical response described by a dielectric function $\varepsilon\left(\omega\right)$.
The reflection amplitudes corresponding to the two polarizations $p={\rm TE},{\rm TM}$
are thus given by the Fresnel formulas for each mirror 
\begin{eqnarray}
r_j^{\rm TE} \left[i\xi,{\bf k}\right] &=& -\frac
{\sqrt{\xi^{2} \varepsilon\left( i\xi \right) + c^{2} {\bf k}^{2}} - c\kappa}
{\sqrt{\xi^{2} \varepsilon\left( i\xi \right) + c^{2} {\bf k}^{2}} + c\kappa} 
\nonumber \\
r_j^{\rm TM}  \left[i\xi,{\bf k}\right] &=& \frac
{\sqrt{\xi^{2} \varepsilon\left( i\xi \right) + c^{2} {\bf k}^{2}} 
- c\kappa\varepsilon \left( i\xi \right)}
{\sqrt{\xi^{2} \varepsilon\left( i\xi \right) + c^{2} {\bf k}^{2}} 
+ c\kappa\varepsilon \left( i\xi \right)}
\label{Fresnel}
\end{eqnarray}
Taken together, relations (\ref{RealMirrors},\ref{Fresnel}) reproduce the Lifshitz 
expression for the Casimir force at zero temperature \cite{Lifshitz56}. 
It is worth stressing again that relations (\ref{RealMirrors}) have a wider domain 
of validity since, as already discussed, they allow one to deal with more general 
scattering amplitudes than (\ref{Fresnel}). 

The optical response of conduction electrons in metals is approximately described
by a plasma model, that is by a dielectric function
\begin{eqnarray}
\varepsilon\left(\omega\right) &=& 1 - \frac{\omega_{\rm P} ^2}{\omega^2}
\end{eqnarray}
A better description is given by the Drude model which accounts for the
relaxation of conduction electrons
\begin{eqnarray}
\varepsilon\left(\omega\right) &=& 1 - 
\frac{\omega_{\rm P} ^2}{\omega\left(\omega+i\gamma\right)}
\end{eqnarray}
Since the ratio $\frac{\gamma}{\omega_{\rm P}}$ is much smaller than unity,
the relaxation parameter $\gamma$ has a significant effect on $\varepsilon$
only at frequencies where the latter is much larger than unity and where,
accordingly, the mirror is nearly perfectly reflecting.
It follows that relaxation has a limited influence on the value of the
Casimir force \cite{Lambrecht00}.

\begin{figure}[htb]
\centerline{\psfig{file=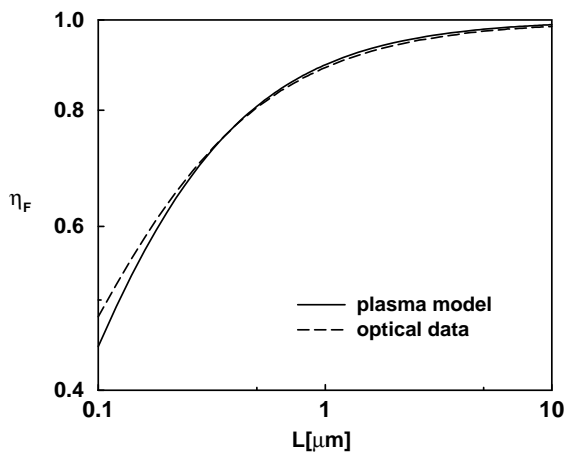,width=8cm}}
\caption{Reduction factor $\eta _{\rm F}$ for the Casimir force between two identical 
Au mirrors at zero temperature as a function of the distance $L$. The solid and dashed 
lines correspond to evaluations based respectively on the plasma model with 
$\lambda_{\rm P}=136$nm and on tabulated optical data \cite{Lambrecht00}.} 
\label{compf} 
\end{figure}

In contrast, the modification of the dielectric constant due to interband
transitions has an observable effect on the Casimir force measured at
distances of the order of the plasma wavelength \cite{Lambrecht00}.
This appears on the results of numerically integrated values of the 
reduction factor $\eta _{\rm F}$ shown on Figure (\ref{compf}). 
The solid line represents the factor calculated for two identical Au 
mirrors described by the plasma model with the plasma wavelength 
$\lambda_{\rm P}=136$nm corresponding to Au. The dashed line represents 
the factor calculated by using the tabulated optical data \cite{Lambrecht00}.

This figure clearly shows that the effect of imperfect reflection is important
at the smallest distances explored in the experiments~: the reduction factor is of
the order of 50\% for Au mirrors at a distance around 0.1$\mu$m. It also appears
that a careful description of the optical properties of metals is necessary
to obtain a precise estimation of the force~: in particular, the description
of metals by the plasma model is not sufficient if an accuracy in the 1\% range
is aimed at.

\section{The effect of temperature}

The preceding estimations were corresponding to experiments 
at zero temperature. But all experiments to date have been performed 
at room temperature and the radiation pressure of thermal field fluctuations 
has a significant contribution to the force at distances larger than or of 
the order of a thermal wavelength \cite{Mehra67,Brown69} 
\begin{equation} 
\lambda _{\rm T} = \frac{\hbar c}{k_{\rm B}T} 
\end{equation}
with $\lambda _{\rm T} \sim 7\mu$m at room temperature.  

It is in principle quite simple to describe the effect of thermal field 
fluctuations which are superimposed to vacuum fluctuations. At zero 
temperature indeed, the field energy per mode is simply the vacuum 
contribution $\frac 12 \hbar \omega$. At a non zero temperature, the
field energy is the sum of this vacuum contribution and of the energy
of the mean number $n$ of photons per mode given by Planck law
\begin{eqnarray}
\frac 12  \hbar \omega &\longrightarrow&  
\left(\frac 12 + n \right) \hbar \omega 
\end{eqnarray}
This means that the contribution of a mode of frequency $\omega$
to the Casimir force has to be multiplied by a factor
\begin{eqnarray}
1 + 2n\left(\omega\right) &=&  
\frac 1 {\tanh\frac{\hbar\omega}{2k_{\rm B}T}}
\end{eqnarray}
After the analytical continuation to the imaginary axis, expression 
(\ref{RealMirrors}) has to be modified by inserting a factor
$1 + 2n\left(i\xi\right)$ in the integrand.
This factor has poles at the Matsubara frequencies 
$\xi_m = m \frac{2\pi k_{\rm B}T}\hbar$ ($m$ integer).
The first of these poles lies at zero frequency where the metallic
response functions also diverge and it must therefore be treated with 
great care. 
This delicate point has recently given rise to a burst of controversial 
results for the evaluation of the Casimir force between real dissipative
mirrors at a non zero temperature \cite{Bostrom00,Svetovoy00,Bordag00}
(see also \cite{Lamoreaux01c,Sernelius01r,Sernelius01c,%
Bordag01r,Klimchitskaya01,Bezerra02,Torgerson02}).

Here we use equation (7) of \cite{Genet00} as the starting point of numerical 
integration of the correction factor $\eta_{\rm F}$.
This equation is based on a uniform expansion of the terms to be integrated 
to obtain the Casimir force and it is valid for all the optical models 
of real mirrors.
As far as the recent controversy is concerned, the evaluations deduced in this 
manner are in agreement with the results of \cite{Bordag00} and at variance
with the conclusions of \cite{Bostrom00,Svetovoy00}. 

\begin{figure}[htb] 
\centerline{\psfig{file=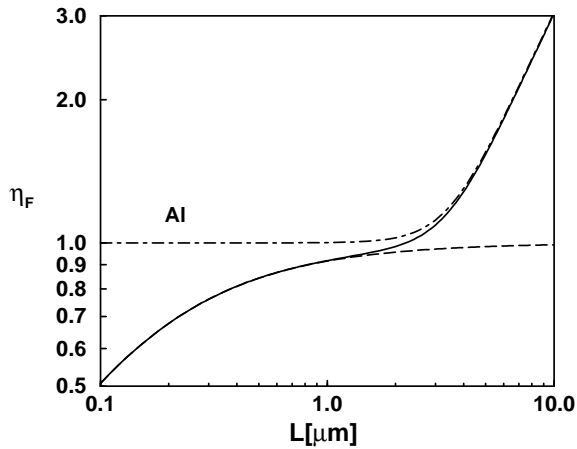,width=8cm}} 
\caption{Correction factors for the Casimir force between two identical Al mirrors 
described by a plasma model with $\lambda_{\rm P}=107$nm at room temperature $T=300$K 
as functions of the distance $L$. The solid, dashed and dotted-dashed lines represent 
respectively the whole correction factor $\eta _{\rm F}$, the plasma correction factor 
$\eta _{\rm F}^{\rm P}$ describing only the effect of imperfect reflection and the 
thermal correction factor $\eta _{\rm F}^{\rm T}$ describing only the effect of 
temperature.} 
\label{etatherm} 
\end{figure}

The resulting correction factor is drawn on Figure \ref{etatherm} as a function of the
distance $L$. Here, we have chosen to consider two identical Al mirrors described by a 
plasma model with the plasma wavelength $\lambda_{\rm P}=107$nm.
The solid line represents the correction factor $\eta_{\rm F}$ in such a configuration 
at room temperature $T=300$K. For the sake of comparison, we have also represented,
as the dashed line, the plasma correction $\eta_{\rm F}^{\rm P}$ evaluated with 
the same mirrors at zero temperature and, as the dotted-dashed line, the thermal
correction $\eta_{\rm F}^{\rm T}$ evaluated with perfect reflectors at room temperature. 

The plasma correction factor $\eta _{\rm F}^{\rm P}$ describes only the effect of 
imperfect reflection and corresponds to the reduction of the force discussed in
the preceding section. Meanwhile the thermal correction factor $\eta _{\rm F}^{\rm T}$ 
describes only the effect of temperature~: it is computed for perfect reflection
and corresponds to an increase of the force. The two factors are appreciable 
respectively at distances smaller than 1$\mu$m and larger than 1$\mu$m. 
It follows that the whole correction $\eta _{\rm F}$ giving the force $F$
when both effects are simultaneously accounted for is nearly equal 
to the product of the plasma and thermal correction factors. 
This is however an approximation the accuracy of which has to be carefully
discussed when a precise evaluation is aimed at.

In order to evaluate the quality of this approximation, it is worth writing
the whole correction factor as
\begin{eqnarray} 
&&\eta _{\rm F} = \eta _{\rm F}^{\rm P} \eta _{\rm F}^{\rm T}
\left( 1 + \delta_{\rm F} \right)   \label{deltaF} 
\end{eqnarray}
A null value for $\delta_{\rm F}$ would mean that the whole correction factor 
may effectively be evaluated as the product of the plasma and thermal corrections
computed independently from each other. In contrast, a non null value 
represents a correlation of the plasma and thermal corrections.        

The correlation factor $\delta_{\rm F}$ has been discussed in
a detailed manner in \cite{Genet00,Genet02}. It should 
be taken into account when an accuracy at or beyond the $1\%$ level is needed. 
This stems from the fact that the correlation scales as the ratio 
$\frac{\lambda_{\rm P}}{\lambda_{\rm T}}$ of the two wavelengths 
which characterize respectively the plasma and thermal effects and is of the order
of $10^{-2}$ for ordinary metals at room temperature. 
The correlation factor is appreciable at distances larger than 1$\mu$m where the 
plasma model is known to be a good effective description of the metallic optical 
response. This justifies the use of this model in \cite{Genet00,Genet02}.
At short distances, say around 0.1-0.5$\mu$m, a more complete description of the metallic 
optical response is needed 
but the temperature correction is negligible in this distance range.
Note also that an analytical approximation of the correlation factor has been given 
in \cite{Genet00} through a perturbative development of the force to first order 
in $\frac{\lambda _{\rm P}}{\lambda _{\rm T}} $. 
The resulting expression is found to fit well the results of the complete numerical 
integration, with an accuracy much better than the $1\%$ level. It provides one with 
a simple method for getting an accurate theoretical expectation of the Casimir force 
throughout the whole distance range explored in the experiments.

\section{Effect of the geometry}

It now remains to describe how the effect of geometry is included
in the theoretical estimations of the Casimir force.

As already discussed, most experiments are performed in a sphere-plane geometry
which differs from the plane-plane geometry for which exact expressions are
available. The force in the former geometry is derived from the Deriagin 
approximation \cite{Deriagin68} which basically amounts to sum up the
contributions corresponding to various inter-plate distances as if 
these contributions were independent. 
In the plane-sphere geometry, the result is simply determined by the
radius $R$ of the sphere and by the Casimir energy as evaluated in the 
plane-plane configuration 
\begin{eqnarray}
F_{\rm sphere-plane} &=& \frac {2\pi R}{A}  E_{\rm plane-plane} \nonumber \\
E_{\rm plane-plane}  &=& \int\limits_{L}^{\infty} {\rm d}x \ F_{\rm plane-plane} 
\left(x\right) = \eta _{\rm E} E_{\rm Cas}  
\end{eqnarray}
We have introduced a correction factor $\eta _{\rm E}$ for the Casimir energy,
evaluated for the plane-plane geometry in the same manner as $\eta _{\rm F}$ 
for the Casimir force in (\ref{defEta}).

\begin{figure}[htb]
\centerline{\psfig{file=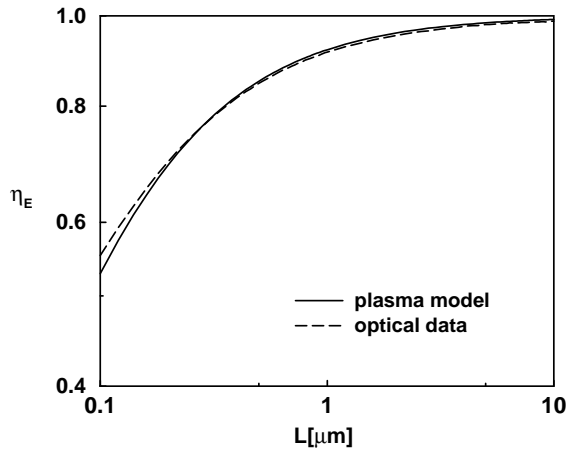,width=8cm}} 
\caption{Reduction factor $\eta _{\rm E}$ for the Casimir energy between two identical
Au mirrors at zero temperature as a function of the distance $L$; same conventions as 
on figure \ref{compf}.} 
\label{compe} 
\end{figure}

Collecting these results leads to the final expression of the Casimir
force in the sphere-plane geometry 
\begin{eqnarray}
F_{\rm sphere-plane} &=& \frac{\hbar c \pi^3 R}{360 L^3}  \eta _{\rm E} 
\label{pft}
\end{eqnarray}
We have shown on Figure (\ref{compe}) the numerically integrated values of the 
reduction factor $\eta _{\rm E}$ for two identical Au mirrors at zero temperature.
As on Figure (\ref{compf}), the solid line represents the factor calculated for 
mirrors described by the plasma model with $\lambda_{\rm P}=136$nm whereas
the dashed line represents the factor deduced from the tabulated optical data for Au
\cite{Lambrecht00}.

We have considered here the case of a null temperature so that the evaluation
is correct only at distances smaller than 1$\mu$m which corresponds to
the most precise results. At longer distances, the temperature correction has 
to be taken into account by following the method presented in the preceding section.

At short distances, surface roughness corrections are also significant. 
They are included by using again the Deriagin approximation
\cite{Klimchitskaya99}, which amounts to average the value of the Casimir 
forces on the various values of the inter-plate distances.
A recent publication \cite{Emig01} opens the route to more precise 
evaluations of the plate corrugation and, potentially, of the 
surface roughness. As it could be expected, the effect of
corrugation is found to depend on the wavelength of the surface
perturbation and not only on its amplitude. In this new evaluation,
the result of the Deriagin approximation is recovered only at the
limit of large wavelengths or, equivalently, small wavevectors
of the surface perturbation.

At this point, it is worth noting that the problem is in fact a more
general deficiency of the Deriagin approximation. This approximation
amounts to add the contributions corresponding to different distances
but we know with certainty that the Casimir force is not additive 
(see a recent detailed discussion in \cite{Barton01}).
As a result, the Deriagin method, though often called the proximity
force theorem, can not be exact. A few results are available for the
plane-sphere geometry which suggest that the approximation leads to
correct results when the radius of the sphere is much larger than the 
distance of closest approach \cite{Langbein71,Keifer78,Balian78}.
But the accuracy of the approximation is not known in a more general
case or for an application to the evaluation of roughness effects. 

\section*{Summary}

It is clear that the Casimir effect has now been unambiguously observed~: 
the experimental precision is already at the 1\% level and it will certainly 
be improved in the future. This precision has allowed the experiments to
observe the effect of imperfect reflection. However, the effect of temperature
has not been seen at the largest distances explored in the experiments 
although it should have been. This is probably due to an insufficient
precision at these distances.

An accurate theory-experiment comparison requires not only precise measurements 
but also accurate and reliable theoretical estimations. 
Important advances have been recently reported for the estimation of the effects 
of imperfect reflection and non null temperature. 
Efforts are still needed for the effects of geometry and surface roughness.
It is worth keeping in mind that not only the accuracy of the 
approximations used to treat these effects should be carefully studied 
for perfect mirrors in vacuum but also that the corrections due to
these effects are probably correlated to the effects of imperfect
reflection and temperature in the same manner as the two latter effects
are now known to be correlated to each other. 

An attractive alternative is to come back to the initial plane-plane geometry
but experiments in this geometry have not been able so far to reach the
precision of sphere-plane experiments.

New advances are expected to occur quite soon in this domain, both on 
the experimental and theoretical sides. These new results will
probably allow one to progress towards an improvement of the
precision of the theory-experiment comparison.
Any such improvement, at the 1\% level or beyond, is important,
since it either confirms a central prediction of Quantum Field Theory or
otherwise reveals surprising new results in the domain of forces with 
nanometric to millimetric ranges. 

\medskip
\noindent
{\bf Acknowledgments}\\
The writing of this report has greatly benefited of discussions and 
correspondance with C. Genet, M.T. Jaekel, G. Barton, F. Capasso, E. Fischbach, 
S. Lamoreaux, J. Long, U. Mohideen, R. Onofrio and C. Speake.

%\def\thefootnote{\fnsymbol{footnote}}
%\vspace{.5cm}

\end{document}